\RequirePackage{fix-cm}
\documentclass[twocolumn]{svjour3}          
\usepackage{graphicx}
\usepackage{tabularx}
\usepackage{enumitem}
\usepackage{multirow}
\graphicspath{{images/}}
\usepackage{float}
\usepackage{wrapfig}
\usepackage[justification=centering]{caption}
\usepackage[T1]{fontenc}
\usepackage{hyperref}
\hypersetup{
    colorlinks=true,
    linkcolor=blue,
    urlcolor=blue,
    citecolor=blue,
}
\begin{document}
\sloppy

\title{ArchiveWeb: Collaboratively Extending and Exploring Web Archive
  Collections} 
\subtitle{How would you like to work with your collections?}


\author{Zeon Trevor Fernando \and Ivana Marenzi \and Wolfgang Nejdl}


\institute{Zeon Trevor Fernando \and Ivana Marenzi \and Wolfgang Nejdl \at
              \email{\{fernando,marenzi,nejdl\}@L3S.de}\\
              L3S Research Center, Hannover, Germany              
}

\date{Received: date / Accepted: date}

\maketitle

\begin{abstract}
  Curated web archive collections contain focused digital content
  which is collected by archiving organizations, groups, and
  individuals to provide a representative sample covering specific
  topics and events to preserve them for future exploration and
  analysis. In this paper, we discuss how to best support
  collaborative construction and exploration of these collections
  through the ArchiveWeb system. ArchiveWeb has been developed using
  an iterative evaluation-driven design-based research approach, with
  considerable user feedback at all stages.

  The first part of this paper describes the important insights we
  gained from our initial requirements engineering phase during the
  first year of the project and the main functionalities of the
  current ArchiveWeb system for searching, constructing, exploring, and
  discussing web archive collections. The second part summarizes the
  feedback we received on this version from archiving organizations
  and libraries, as well as our corresponding plans for improving and
  extending the system for the next release.  
\end{abstract}

\section{Introduction}\label{sec:Introduction}

The web is becoming an important corpus for studying human society by
researchers in the humanities, social sciences, and computer sciences
alike. Web archives collect, preserve, and provide ongoing access to
ephemeral web pages and hence encode important traces of human
thought, activity, and history.  Curated web archive collections
contain focused digital content from archiving organizations, groups, 
and individuals related to specific topics or covering specific
events, which are collected to provide representative samples and
preserve them for future exploration and analysis. However, there have
been only a few concerted efforts to provide tools and platforms for
exploring and working with such web archives.


This article focuses on the
ArchiveWeb\footnote{\url{http://archiveweb.l3s.uni-hannover.de/aw/index.jsf}}
platform which provides facilities to collaboratively explore and work
with web collections in an interactive and user-friendly way, both for
research and for learning. In particular we focus on supporting {\em
  collaborative exploration and analytics\/} through a user friendly
searching and sharing interface (ArchiveWeb). Our goal is to allow
users (e.g., archivists and librarians, web archive curators,
researchers) to pose queries in the context of a larger exploration
process where search results are stored for sharing with colleagues
and collaborators, later discussion, and analysis~\cite{tlt12}. This
enables the user to consider dependencies between queries independent
of the order in which they are posed, thereby regarding them in the
context of later queries and query results. Building on previous work
we did for collaborative learning environments~\cite{tlt12}, we enable
users to build their collections incrementally and collaboratively,
which is an important step to support their research work. We also
allow researchers and their collaborators to approve, add, or remove
results to their collections and to discuss their significance.

In the following section, we provide a short overview of related work,
in Sect. 3, we describe a life cycle model for web archiving, 
and in Sect. \ref{sec:User Requirements}, we give more details about
the analysis of user requirements on which we based the ArchiveWeb
interface. In Sect. \ref{sec:Web Archive Collections}, we describe
two sample web archive collections we are working with. Section 
\ref{sec:System} includes a description of the main functionalities of
our system, and Sect. \ref{sec:Evaluation} presents the evaluation
design with a detailed discussion of the evaluation results. In
Sect. \ref{sec:Future Improvements}, we summarize future extensions
for the next ArchiveWeb release and finally draw conclusions in
Sect. \ref{sec:Conclusion}.

\section{Related Work}\label{sec:Related Work}
Information retrieval (IR) models and algorithms have been extremely
successful during the last 20 years in providing everybody with easy
access to the vast amount of information available on and through the
web. Web and IR conferences connect a large number of researchers
working on problems related to web search, the Social Web and data
mining, and related topics. Surprisingly little work has been spent,
however, on issues related to temporal retrieval, or exploration and
analysis of large temporal collections like web
archives~\cite{dougherty09}, even though the need for focusing
research on these issues has been recognized~\cite{twaw11,wtst11}.

The Alexandria project\footnote{\url{http://alexandria-project.eu/}}
focuses on open questions where radical changes and advances in the
state of the art are necessary to move ahead in our abilities of
indexing, retrieving, and exploring past web content. Models and
algorithms for temporal information retrieval developed in the context
of this project will take the unique temporal dimension of web
archives into account. Semantic entity-based indexing will support
exploration of temporal web content and evolving entities in a more
user-oriented way than conventional document-based retrieval. Finally,
user input gathered by complex and collaborative search and analysis
processes of archive users will further enable us to improve web
archive indexing and enrichment considerably. While this project
targets web archives, similar issues arise in any digital archive, and
solutions developed within Alexandria will likely be applicable to
other types of archives as well.

Often desired search functionalities in web archives include full-text
search with good ranking, followed by URL search~\cite{archiving07}. A
recent survey showed that 89\% of web archives provide URL search
access, and 79\% give metadata search
functionalities~\cite{gomes11}. Some existing projects that provide
limited support for web archive research are discussed below.

The Wayback Machine\footnote{\url{http://archive.org/web/}} is a web
archive access tool supported by the Internet Archive. It provides the
ability to retrieve and access web pages stored in a web archive
through URL search. The results for each URL are displayed in a
calendar view which displays the number of times the URL was crawled
by the Internet Archive web crawlers. Archive-It and
ArchiveTheNet\footnote{\url{http://archivethe.net/}} are web archive
services provided by the Internet Archive and the Internet Memory
Foundation. These services enable focused archiving of web contents by
organizations, such as universities or libraries, that otherwise could
not manage their own archives.  The Memento
Project\footnote{\url{http://timetravel.mementoweb.org}} enables the
discovery of archived content from across multiple web archives via
URL search.

A few researchers have worked on providing new interfaces and
visualizations for searching, exploring, and discovering insights from
web archives. Odijk et al.~\cite{odijk15} present an exploratory
search interface to improve accessibility of digital archived
collections for humanities scholars, in order to highlight different
perspectives across heterogeneous historical collections. The
motivation for this work derives from the huge amount of digital
material that has become available to study our recent history,
including books, newspapers, and web pages, all of which provide
different perspectives on people, places, and events over time. The authors connect heterogeneous digital collections
through the temporal references found in the documents as well as
their textual content, in order to support scholars to detect,
visualize, and explore materials from different perspectives. Padia
et al.~\cite{padia12} provided an overview of a web archive collection
by highlighting the collection's underlying characteristics using
different visualizations of image plots, wordle, bubble charts, and
timelines. Lin et al.~\cite{lin14} present an interactive
visualization based on topic models for exploring archived content, so
that users can get an overview of the collection content. The
visualization displays a person-by-topic matrix that shows the
association between US senators' websites and the derived topics. The
interface also provides drill-down capabilities for users to examine
the pages in which a topic is prevalent.

All of the above tools and interfaces help support the exploration and
search of web archives for individual users and researchers. In
addition, ArchiveWeb aims at supporting the collaborative exploration
of web archives.  Previous research on helping users keep track of
their resources includes tools that provide better search and
organizational facilities based on metadata/time~\cite{dumais03} or
tagging~\cite{cutrell06}. Our system provides similar organizational
functionalities refined through close collaboration with several
learning communities and previous work on the LearnWeb
platform~\cite{tlt12}, thus gaining advantage from several years of
development and user feedback in that context. ArchiveWeb builds on
the LearnWeb experience which already supports collaborative
sensemaking by allowing users to share and collaboratively work on
resources retrieved from various web sources.

LearnWeb\footnote{\url{http://learnweb.l3s.uni-hannover.de}} is a
learning and competence development environment, which allows users to
share and collaboratively work on resources collected from the web or
user-generated~\cite{tlt12}. It provides users with a search interface
for resource discovery and sharing across various Web 2.0 services
such as YouTube, Flickr, and Slideshare, including LearnWeb itself,
and offers a Personal Web 2.0 Learning Space. In order to support
collaborative searching, LearnWeb provides automatic resource
annotation. Resources in LearnWeb can be bookmarked, tagged, rated,
and discussed by all users who are allowed to access them. Comments on
particular learning resources can be used to enrich the
description. Users can create folders to bundle resources that belong
to the same learning context. Hence, the LearnWeb community can
collaboratively identify the best learning resources for specific
learning domains.  A discussion of the full potentialities and
affordances of LearnWeb as a collaborative platform is beyond the
scope of this paper, but can be found in a series of published studies~\cite{marenzi14,marenzi12,tlt12}.

The Integrated Digital Event Archive and Library (IDEAL)
system\footnote{\url{http://www.nsf.gov/awardsearch/showAward?AWD_ID=1319578}}
addresses the integration of digital libraries and archival
technologies in support of stakeholders interested in studying
important events. It focuses on events falling into two broad
categories: (1) related to crises or tragedies as well as recovery and 
(2) government/community-related events (e.g., elections,
demonstrations). The system monitors web-based (news, government, and
other websites) and social media activity (tweets) to automatically
detect interesting events and respond to general and specific requests
archiving requests as well. Once events are identified, intelligent
focused crawling and filtering approaches are used to ingest content and 
generate collections with high precision, recall, and low bias as
is needed for scholarly study by social scientists. 
The system also incorporates a wide range of integrated services such as browsing,
searching, recommendation, clustering, sentiment analysis,
summarization and visualization of data, information, or context.

The Big UK Domain Data for the Arts and Humanities (BUDDAH)
project\footnote{\url{http://buddah.projects.history.ac.uk/}} aimed to
develop a theoretical and methodological framework within which to
study the archived UK web and web archives in
general~\cite{jane17}. To demonstrate the value of web archives to
humanities researchers and to build a community, a wide range of
research projects was undertaken - from analyzing disability action
groups online to studying the Ministry of Defense's recruitment
strategy. The project also worked on developing a suite of tools to
support the analysis of the UK web archive and introduced an
exploratory search interface
SHINE\footnote{\url{https://www.webarchive.org.uk/shine}} through
which humanities scholars and social scientists accessed web
archives~\cite{jackson16}.

The RESAW network,\protect\footnote{\url{http://resaw.eu/about/}}
established in 2012, aims to establish and operate a collaborative
world-class transnational European research infrastructure to enable
cross-border studies of the archived web by integrating and opening up
existing national web archives. It will facilitate easy access to
large amounts of cultural heritage big data, as well as provide
searching, selecting, and analysis of the material itself, thus making
the research process more efficient and enhancing the European
research area. At the moment, RESAW is a network of scholars and
libraries interested in Web Archives and a research proposal to which
we are contributing.  We intend to make ArchiveWeb available to the
participating libraries, both to get additional requirements from our
partners as well as to support the collection building process for
various national web archives.

\section{ The Web Archiving Life Cycle Model}\label{sec: The Web Archiving Life Cycle Model}

\begin{figure}[!ht]
\centering
\includegraphics[width=\columnwidth]{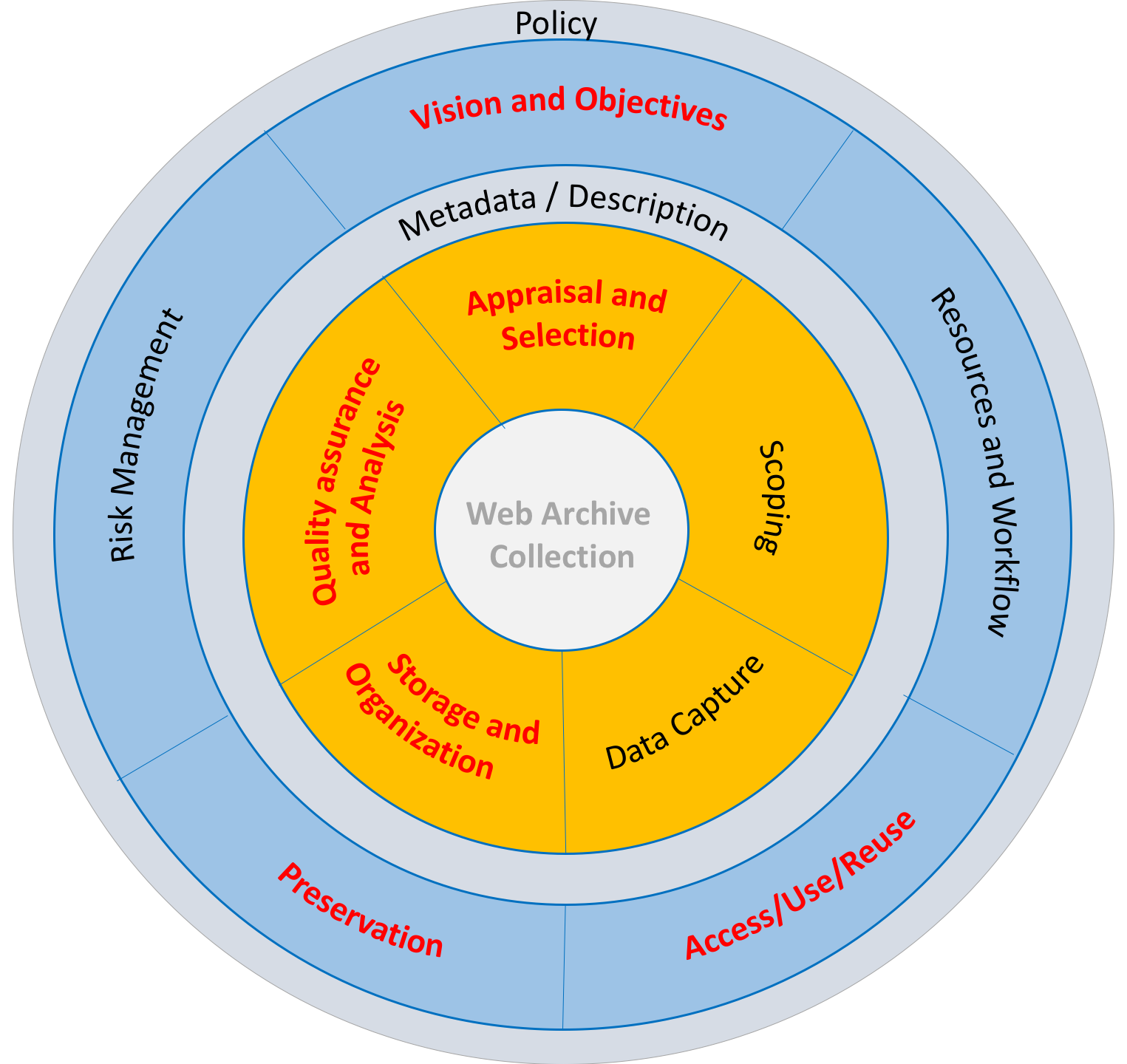}
\caption{Web Archiving Life Cycle Model}
\label{walcm}
\end{figure}

Despite the increasing number of web archiving programs, best
practices and a common methodology for web archiving have not been
established yet~\cite{walcm13}. The Web Archiving Life Cycle Model
(WALCM), developed by experts at the Internet Archive, suggests a
common framework for organizations seeking to archive the web. The
model, reproduced in Fig. \ref{walcm}, includes several phases
representing the various steps of a common workflow which
organizations can refer to in order to create or improve their web
archiving programs.

The original model is focused on the institutional policy collection;
target users are archiving institutions and curators of such
collections. Each phase of this process is supported by the leading
web archiving service Archive-It (mentioned above).

In the first phase of our work, we referred to the original WALCM
model to prepare the investigation of current practices and of user
requirements (Sect. \ref{sec:User Requirements}). Figure \ref{walcm}
shows the steps of the WALCM model highlighted in red which we used as
a reference to guide the semi-structured interviews and to analyze the
results. We focused on three steps related to high-level policy
decisions (i.e., Vision and Objectives, Access/Use/Reuse,
Preservation), and on three day-to-day tasks of the web archiving
process (i.e., Appraisal and Selection, Storage and Organization,
Quality Assurance and Analysis).

As regards the curatorial activities of storage and organization, and
quality assurance and analysis, our interpretation is slightly
different from the original one, though. With storage and organization
we do not refer to the temporary or long-term storage plan of the
organization, but rather we consider the possibility for ArchiveWeb users
to organize resources in sub-collections. Quality assurance and
analysis is based on the possibility for ArchiveWeb users to
collaboratively annotate and comment on resources in order to state or
to discuss their relevance.

\section{User Requirements Analysis 2015}\label{sec:User Requirements}

\subsection{Motivation and Setup}

The motivation of our study was to provide a user-friendly interface
to allow various kinds of users to access archived web collections and
use the resources in various scenarios, e.g., for research or for
learning.

In order to do so, in Summer 2015 we carried out a preliminary
analysis of user requirements~\cite{Fernando2016}. Participants were
colleagues from the Internet Archive, as well as from libraries and
similar institutions who are creating/curating Archive-It collections,
e.g., Stanford University Libraries, Cornell University, University of
Toronto, Columbia University, the National Museum of Women in the
Arts, and the New York Arts Resources Consortium.  The purpose of the
study was to collect and analyze requirements for building and using
web archive collections to be used as important input for new
algorithms and tools to improve web archives.  The ultimate goals of
our work are to
\begin{itemize}
\item understand what kinds of web archive collections are being
  curated by different kinds of organizations 
\item contribute further improvements for working with web archive
  collections, especially regarding interface and search
  functionalities
\end{itemize}

With reference to the web archiving life cycle (Fig. \ref{walcm}), our
interest was focused on specific steps related to policy decisions
(e.g., Vision and Objectives, Access/Use/Reuse, Preservation), and
data management (e.g., Appraisal and Selection, Storage and
Organization, Quality Assurance and Analysis). Based on these six main
categories, we used five leading questions to guide the
semi-structured interviews:
\begin{enumerate}
\item What collections are built and what is the
  motivation for building these collections?
\item Who decides which collection is built and who is responsible for selecting which resources go into a specific collection?
\item Who are/could be potential users of a collection? Do
  you track the use of your collection?
\item Which functionalities are important for users who work
  with these collections when accessing Web Archive collections?
\item Are there any user interface improvements which you and
  your users already have on their wish list?
\end{enumerate}
Respondents were free to add information and details related to their
specific role/experience in the archiving process.

Interviews were conducted between August 3 and 6, 2015, 
during a research visit of the investigators in San Francisco. We
interviewed eight experts playing different roles and taking care of
various activities in the archiving process, including the director of
the Web archiving program at the Internet Archive, three Web archiving
coordinators of University libraries, two Web archiving coordinators
in national museums, one digital archivist, and one head of metadata
services. The experts were selected based on suggestions provided by the Internet Archive. 
Three interviews were conducted in person at the Internet
Archive premises; the remaining five were scheduled online according
to the availability of the experts. Interviews took on average half an
hour each, and the discussions were recorded for data analysis
purposes. We used a consent form to inform the participants on the
procedures and protection precautions and to collect their signatures
for agreement.

Participation in the project was voluntary, and anybody was free to
withdraw from the project at any point. Personal information is
treated confidentially: Name and affiliations are anonymized in any
resulting publications, unless participants give us explicit consent
to identify them as a subject.

All recordings have been transcribed by one of the investigators, and
qualitative content analysis of the transcripts has been carried out
manually by two coders. Answers have been summarized referring to the
categories in the original WALCM model (Fig. \ref{walcm}). In Table
\ref{table:1}, we report a summary of the answers referring to the
categories in the original WALCM model.

\begin{table*}[htbp]
\centering
\begin{tabularx}{\textwidth}{|>{\raggedright}p{5cm}|p{11.5cm}|}
\hline
Theme & Answers \\
\hline
Vision and Objectives 

\textit{What collections are built and what is the motivation for building
these collections?} & 
In most cases the motivation for starting the archiving of digital
collections has been to preserve: 
\begin{itemize}
\item institutional or government websites
\item special collections received as donations of materials from
  individuals or organizations
\item event-based collections (e.g., Olympic games or election campaigns)
\item research outputs (e.g., faculty members hosting their funded
  projects including related assets such as documents and research
  datasets)
\end{itemize} 
Larger institutions have a more structured approach including, in some cases, focused groups with their staff and researchers who determine which areas are in focus with respect to priorities. For smaller institutions starting to archive online content was also seen as a challenge \\
\hline Access/Use/Reuse

\textit{Who are/could be potential users of a collection? Do you track the use
of your collection?} & 
In most cases users are members of the library or of the university
(e.g., professors or researchers). 

For specific collections, most of the requests to access archived web
collections come from people who are looking for documents that have
been removed from a government website.

Very few institutions and libraries track what users do using basic
metrics (e.g., number of hits on a collection level). \\ 
\hline
Preservation

\textit{What kind of websites should be preserved?} &
As regards to preservation, the priorities are to:
\begin{itemize}[leftmargin=*]
\item collect non-governmental organizations websites as opposed to
  the government websites 
\item preserve resources which are in danger of disappearing from
  the web 
\end{itemize} \\
\hline
Appraisal and Selection

\textit{Who decides which collection is built and who is responsible for
selecting which resources go into a specific collection?} & 
All respondents confirmed that the current efforts at their
institutions are mainly individual curators or subject specialists who
are responsible for individual collections or individual archiving
requests 
\begin{itemize}[leftmargin=*]
\item in some cases a collection development executive group is appointed
\item few institutions are collecting nominations from experts,
  researchers, library and curatorial staff
\end{itemize}
Digital collections at larger institutions are growing incrementally,
starting for example from collecting the institutional memory, and
including websites of single departments and faculties, student
organization websites, or special collections on specific areas of
interest. Institutions dealing with art (e.g., museums and galleries)
are particularly concerned with preserving resources which are in
danger of disappearing from the web. \\ 
\hline Storage and Organization

\textit{Which functionalities are important for users who work with these
collections when accessing Web Archive collections?} & 
\underline{For curators}

Collaboration is an important aspect in order to avoid duplicating
resources and efforts. Some curators are trying to find ways of
collaborating in terms of 
\begin{itemize}
\item collecting and displaying information 
\item creating and sharing metadata schema
\item sharing folksonomies and tagging
\end{itemize}
Other useful functionalities would be:
\begin{itemize} 
\item limiting the content to a certain number of seeds 
\item downloading metadata
\item providing solutions to archive dynamic web pages
\end{itemize}
\underline{For final users}
\begin{itemize}
\item Fulltext search
\item Annotation
\item Presentational functionalities (e.g., topic modeling, thumbnails
  of the major versions of the homepage per seed, link labels analysis 
\end{itemize} \\
\hline
Quality Assurances and Analysis

\textit{Are there any user interface improvements which you and your users
already have on their wish list?} & 
Improvement of the interface
\begin{itemize}[leftmargin=*]
\item Longitudinal data exploration
\item query of subcollections
\item integration of web archive resources into the general digital
  environment  
\end{itemize} \\
\hline
\end{tabularx}  
\caption{User Requirements Analysis - Summary of responses}
\label{table:1}
\end{table*}

On the basis of this preliminary study, and building on previous work
we did for collaborative learning environments~\cite{tlt12}, we
designed and built the ArchiveWeb system to support collaborative
creation and enrichment of web archive collections, with a focus on
user interface and searching/sharing functionalities. We ingested 200
web archive collections from Archive-It, and asked our archiving
partners for their input. In November 2015, we prepared a Memorandum of
Understanding (MoU) to describe how we integrate/use collections in
ArchiveWeb, and we shared it with all participating institutions to
collect their official agreement and signatures.

In the following, we briefly talk about some example collections, and in the next two sections, we describe
the main ArchiveWeb functionalities and the evaluation provided by the
experts who participated in the user survey.

\section{Web Archive Collections}\label{sec:Web Archive Collections}


Archive-It is a subscription web archiving service from the Internet
Archive that helps organizations to harvest, build, and preserve
collections of digital
content.\footnote{\url{https://archive-it.org/learn-more/}} It was
first deployed in 2006 and is widely used as a service to collect,
catalog, and manage collections of archived web content. Full-text
search is also available, even though effective ranking is still an
open issue.  All content is hosted and stored at the Internet Archive
data centers.

Currently, the Archive-It system is mainly used by librarians and
curators in order to build their collections. Less support is given to
users and domain experts who actually want to work with the
collections, or to the general public. ArchiveWeb supports searching,
collecting, exploring, and discussing web archive collections such as
the ones provided through the web archiving service of the Internet
Archive.

Currently, about 200 collections from Archive-It have been integrated
into the ArchiveWeb system with full metadata indexing, and can be
explored through a visual interface. The Archive-It collections that
were selected consist of those curated by organizations of the experts
we interviewed as part of our user requirements analysis and those
that cover a diverse range of topics such as Humanities, Arts,
Society, Culture and Global Events.

In order to give a representative example of the diversity of the
archived materials, we describe two specific collections.

\subsection{Human Rights Web Archive}

\begingroup

\setlength{\intextsep}{3pt}
\setlength{\columnsep}{3pt}
\begin{wrapfigure}{l}{0.2\textwidth}
\includegraphics[width=\linewidth]{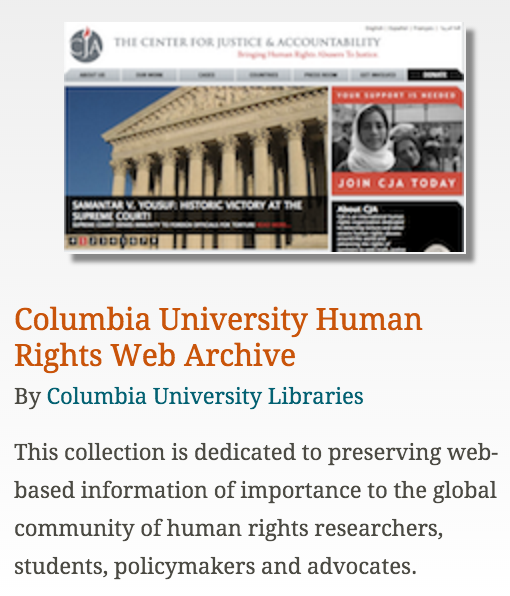} 
\label{hrwa}
\end{wrapfigure}

The collection \textit{Human Rights Web
  Archive}\footnote{\url{https://hrwa.cul.columbia.edu/about}} (HRWA)
by \textit{Columbia University
  Libraries}
is made up of searchable archived copies of websites related to human
rights created by various non-governmental organizations, national
human rights institutions, tribunals, and individuals. The collection
was started in 2008 and is still being continued, adding new websites
on a regular basis.

Identification of websites for archiving is done by subject
specialists with expertise in human rights and different regions of
the world at Columbia University Libraries. Public nominations are
provided by human rights researchers, advocates, organizations, and
individuals who are involved in the creation of human rights-related
websites. Priority is given to websites hosted in countries that do
not have any systematic web archiving initiatives in place. Websites
of intergovernmental organizations such as the United Nations are not
included in the collection.

Archive-It services are used to maintain the
collection, and the Internet Archive and Columbia University Libraries
store copies of the resulting data. The collection includes over \textit{711
  websites} with more than 50 million searchable documents and over
\textit{115 million archived documents} with an archived data size of
more than \textit{5TB}.

\endgroup

The HRWA collection provides a good balance between websites of large
and well-known human rights organizations based in North America and Europe 
and websites of smaller organizations from other regions. 
One example of a website that no longer has a live version but can still be accessed by
researchers via the Web archive collection is TibetInfoNet, which
monitored the situation in Tibet \url{http://www.tibetinfonet.net/}. This
page has been captured between May 2008 and July 2015. 

\begin{figure*}[!ht]
\centering
\includegraphics[width=\textwidth]{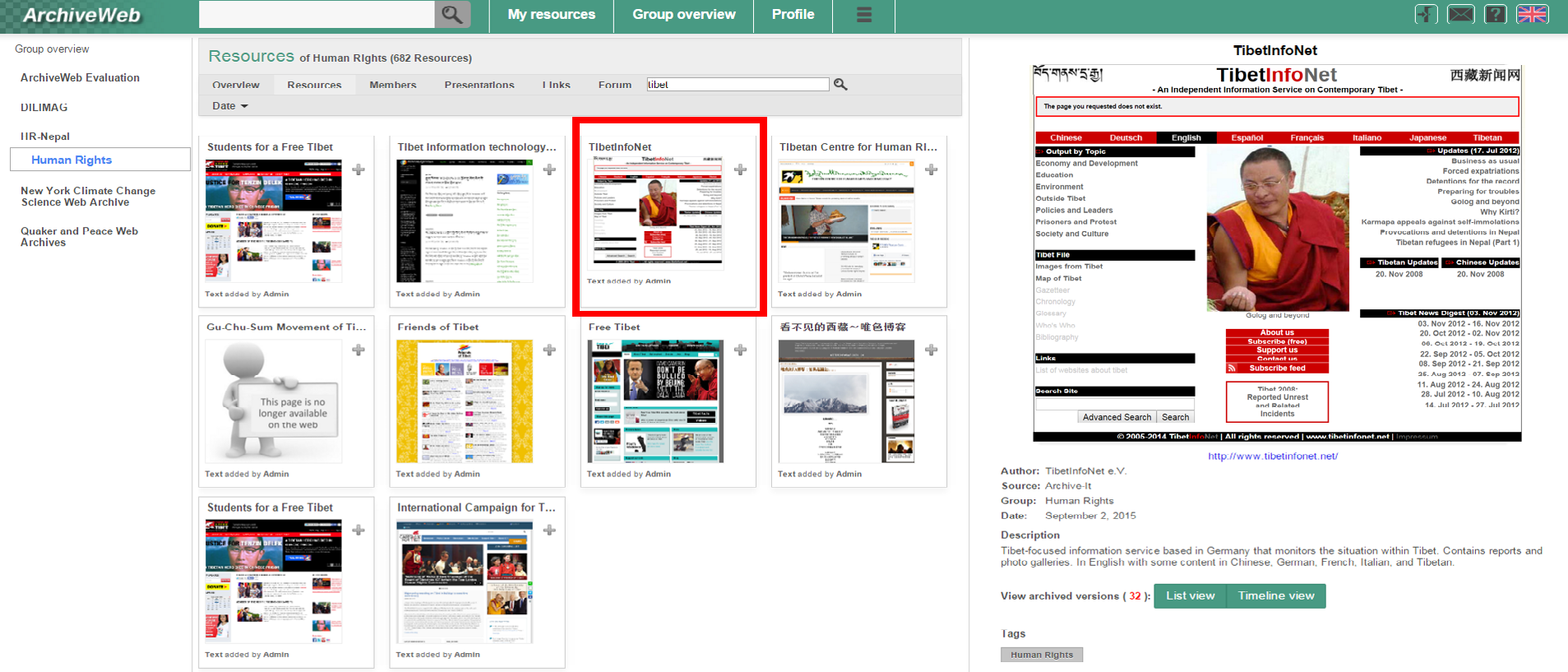}
\caption{Human Rights Example Resource}
\label{hrwa_ex}
\end{figure*}

The ArchiveWeb screenshot shows the human rights group restricted to ``Tibet''
-related resources, with the TibetInfoNet resource on the right side
(Fig. \ref{hrwa_ex}). The URL provided in ArchiveWeb below the screenshot on the right side
points to the original website, which is no longer available on the
live web, but can be retrieved in its previous versions through the
Wayback Machine (as described in Sect. \ref{sec:enriching}).

\subsection{Contemporary Women Artists on the Web}

\begin{figure}[!ht]
\centering
\includegraphics[width=\columnwidth]{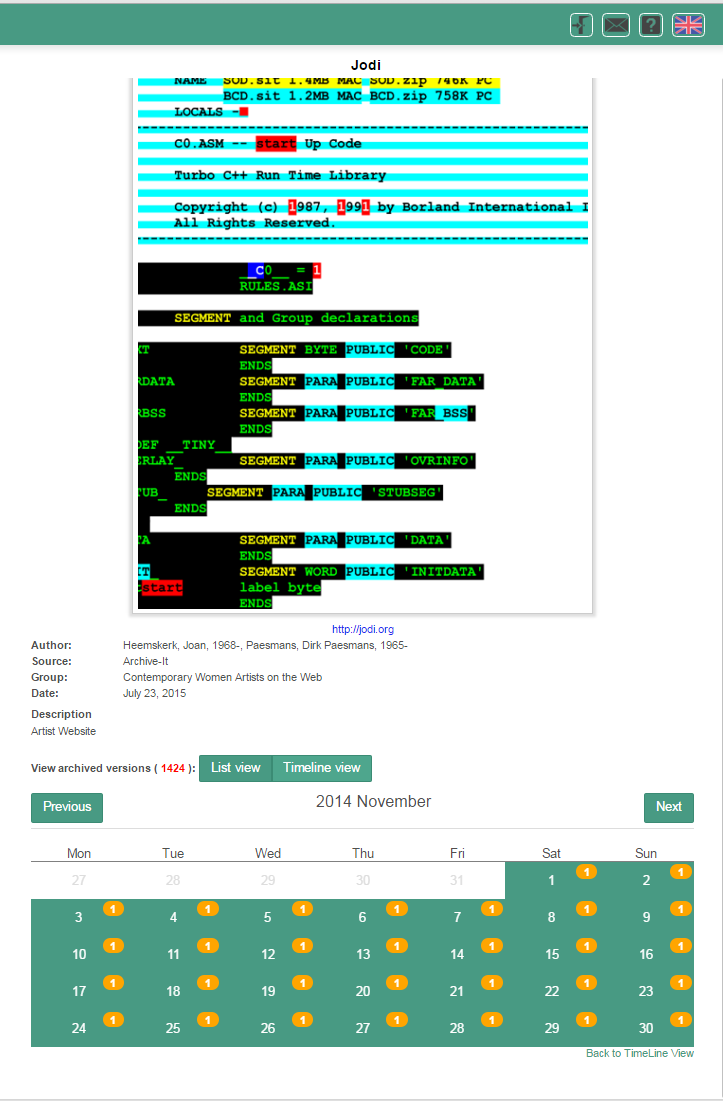}
\caption{Jodi Art Collective}
\label{jodi}
\end{figure}

The collection {\em Contemporary Women Artists on the
  Web}\footnote{\url{https://archive-it.org/collections/2973}} has
been collected by the {\em National Museum of Women in the
  Arts} (NMWA) since December
2011. The NMWA is a gender-specific museum, located in Washington, and
is the only museum solely dedicated to celebrating women's
achievements in the visual, performing, and literary arts. Since
opening its doors in 1987, the museum has acquired a collection of
more than 4,500 paintings, sculptures, and works on paper and decorative
art.

The collection {\em Contemporary Women Artists on the Web} includes three
components: (1) Individual websites created by artists working in
conceptual or new media art, (2) Artist profiles of women artists
represented by contemporary art galleries, and (3) Women artist
organizations.

One representative example of a resource in this collection is
\url{http://jodi.org}, see Fig. \ref{jodi}, which has been captured
1418 times between April 16, 2009, and March 31, 2016, and which includes
726 videos.

As described in
Wikipedia,\footnote{\url{https://en.wikipedia.org/wiki/Jodi_(art_collective)}}
it is the result of the initiative of two authors who work together
toward a shared aim: Joan Heemskerk (born in 1968 in Kaatsheuvel, the
Netherlands) and Dirk Paesmans (born in 1965 in Brussels, Belgium). Their
background is in photography and video art. Since the mid-1990s, they
started to create original artworks for the World Wide Web, and a few
years later, they turned to software art and began the practice of
modifying old video games to create a new set of art
games~\cite{stalker05}. 

Opening windows cascade across the screen, error messages squawk, and
files replicate themselves endlessly. As graphics explode across the
screen, the viewer gradually realizes that what had initially appeared
to be a computer glitch is in reality the work of an irrational,
playful, or crazed human~\cite{lieser09}. Webpages are interactive, and
a new perspective is displayed as the user clicks on various elements
on the page. 

Each day the webpage \url{http://jodi.org} changes, and a new piece of art is
displayed. If the resource was not archived, all work would be
lost. For this reason, it is important to archive at least a capture
once a day in order to document the work of these artists over time.

\section{ArchiveWeb System: Main Functionalities}\label{sec:System}

\begin{figure}[!ht]
\centering
\includegraphics[width=\columnwidth]{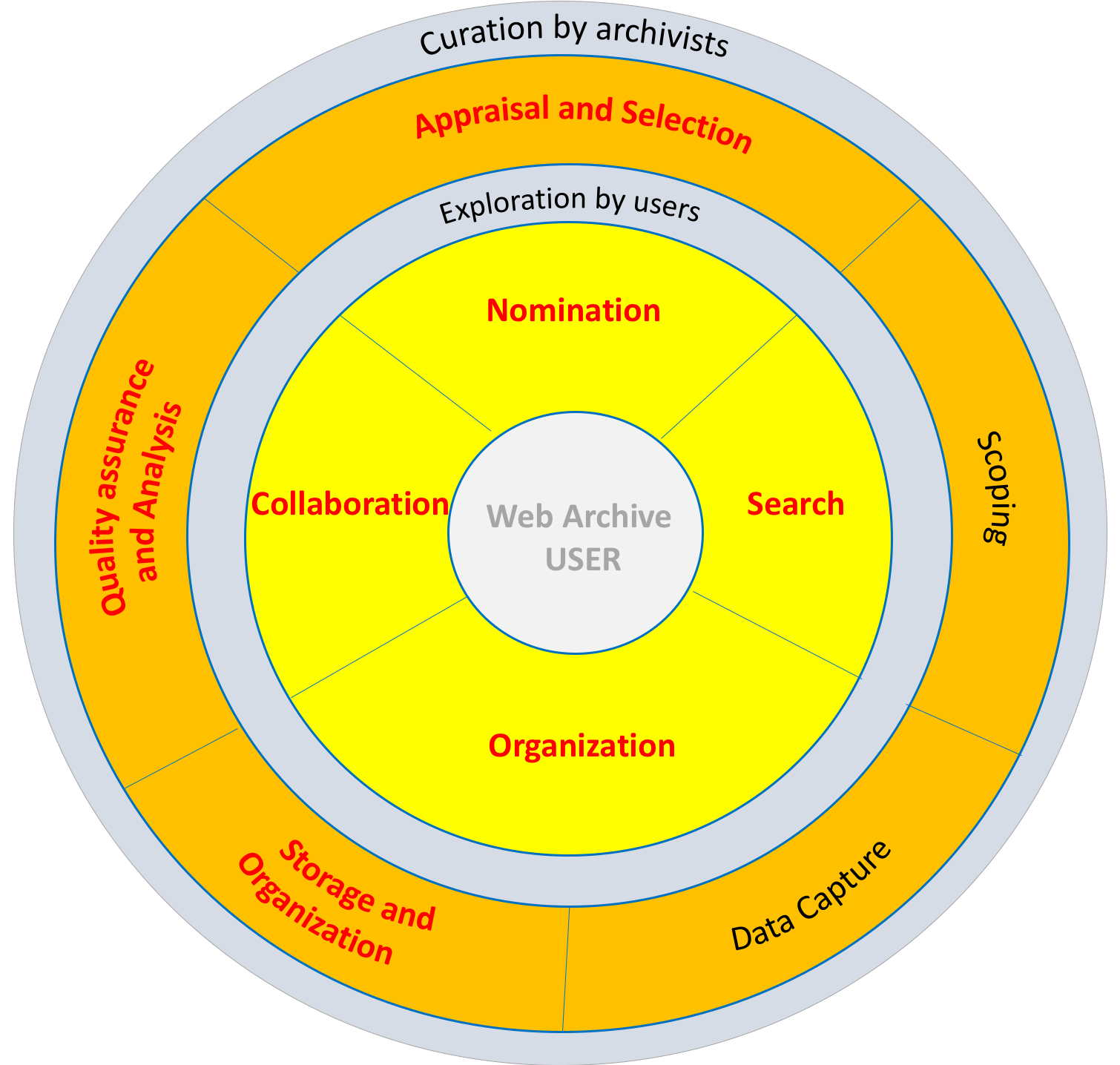}
\caption{The ArchiveWeb User Model}
\label{aum}
\end{figure}

With the requirements described in Sect. \ref{sec:User Requirements}
and the collections sketched in Sect. \ref{sec:Web Archive
  Collections} in mind, we worked on providing the ArchiveWeb system
to web archivists and libraries to facilitate collaborative
exploration of multiple focused web archive collections. 

At the design level, in ArchiveWeb we did not want to replicate
services offered by the Archive-It service such as scoping, data
capture, or risk management. Rather than focusing on the institution
and collection level, we focus more on the final user who wants to
explore and work with web archive collections. Thus, Fig. \ref{aum}
highlights a more detailed inner circle with focus on the Web Archive
User (instead of on the collections). Besides the curatorial tasks
already present in the original WALCM model (i.e., Appraisal and
Selection, Storage and Organization, Quality Assurance and Analysis)
we detail four activities related to collaborative search and
exploration of resources, as well as organizational functionalities
(i.e., Nomination, Search, Organization, and Collaboration).

We used these categories, highlighted in red in the model, to guide
the design of the system functionalities, as well as the evaluation of
the system in 2016 (Sect. \ref{sec:Evaluation}).

The main features provided by ArchiveWeb to support this exploration
are: \textit{search} - searching across multiple collections in
conjunction with the live web, \textit{organization} - grouping of
resources for creating, merging or expanding collections from existing
ones and \textit{collaboration} - enrichment of resources in existing
collections using comments and tags for the purpose of discussion. The
following sections give a detailed description of these main system
functionalities. In Sect. \ref{technical workflow} we describe the
technical workflow of the system, first describing how the data were
collected and then how the system was implemented.

\subsection{Technical Workflow}\label{technical workflow}

\paragraph{Data Collection Process.}The Archive-It collection data was obtained by scraping the Archive-It
website\footnote{\url{https://archive-it.org/explore/?show=Collections}}. We
selected a subset of collections to be crawled based on the subject
metadata field that included the topics: Society \& Culture, Arts \&
Humanities, Universities \& Libraries. Also collections specifically
curated by the evaluators institutions were included in the
crawl.

For each Archive-It collection, we collect and store the following
metadata: collection title, collecting institution, description,
subject, and collectors. All results/seed URLs displayed on each
Archive-It collection page (e.g., Human Rights
collection\footnote{\url{https://archive-it.org/collections/1475}})
are also crawled. For each result, we store the following metadata:
URL, title, description, subject, collector, creator, publisher,
language, format, and type. If a result does not have metadata for title,
subject, or language in Archive-It, we fetch the title, language, and
subject from the meta-keywords if available from the HTML page of the
result. While crawling each result of a collection, we store the
individual capture information fetched using the TimeMap component of
the Memento framework. A TimeMap is a list of URIs of captures of the
original result/URL (e.g.,
4genderjustice.org\footnote{\url{https://wayback.archive-it.org/1475/timemap/link/http://4genderjustice.org/}}).

We update the Archive-It collections to keep ArchiveWeb content up to
date with the information in the Archive-It portal by crawling and
updating only those collections that have any result that has been
captured/crawled recently (3 months before current time of crawl).

\paragraph{Implementation.}The interface is implemented using the MVC
framework JavaServer Faces (JSF) and PrimeFaces UI component library
for JSF. Navigation across the ArchiveWeb system is facilitated
through a grid layout in order to guide and keep the user in context
and engaged at all times. For more details about the implementation
and conceptual design of the initial platform LearnWeb on which the
ArchiveWeb system has been built, we refer the reader to~\cite{tlt12}.

In the following, we discuss the extensions that we incorporated
during the first year of the project in order to support the main
features of ArchiveWeb. To provide access to the crawled Archive-It
data from the above-mentioned process, they have to be modeled as
digital objects (resources, groups/collections) and stored in a
relational database format, using the MariaDB repository. The data in
the collections are made searchable through a metadata index, using
Solr. Comments and tags added to resources by various collaborators
while curating collections are also indexed in order to support
collaborative search and exploration. Search filters based on the
various metadata fields of the results were incorporated using the
Solr faceted search feature to narrow down search results.

As each of the resources/URLs has a number of archived versions, the
challenge was how to incorporate navigation through these versions in
the interface. To support users to navigate through the different
versions, a \textit{timeline} view similar to the Wayback Machine
interface was incorporated (see Fig. \ref{group}). A simple
\textit{list} view was also included to provide a one-step navigation
for resources with very few archived versions. Finally, we also
implemented a functionality for merging existing collections and
creating new ones, so that users have the flexibility to merge
multiple collections covering similar topics/subjects and build upon
them.

\subsection{Search}\label{sec:search} 

\begin{figure}[!ht]
\centering
\includegraphics[width=\columnwidth]{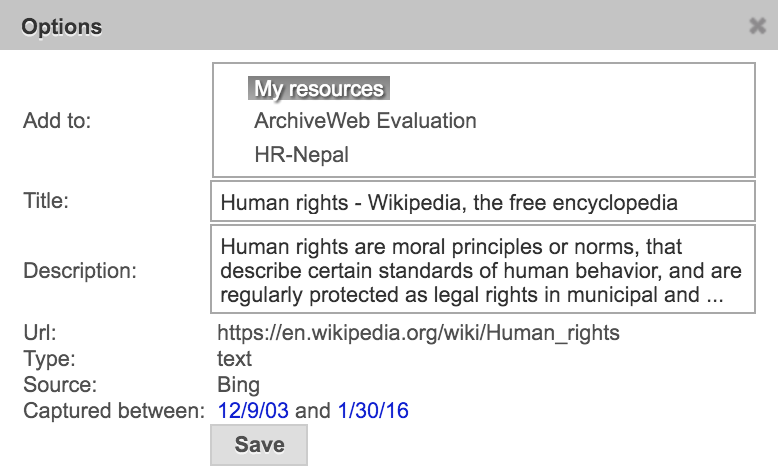}
\caption{Options Dialog}
\label{search options}
\end{figure}

\begin{figure*}[!ht]
\centering
\includegraphics[width=\textwidth]{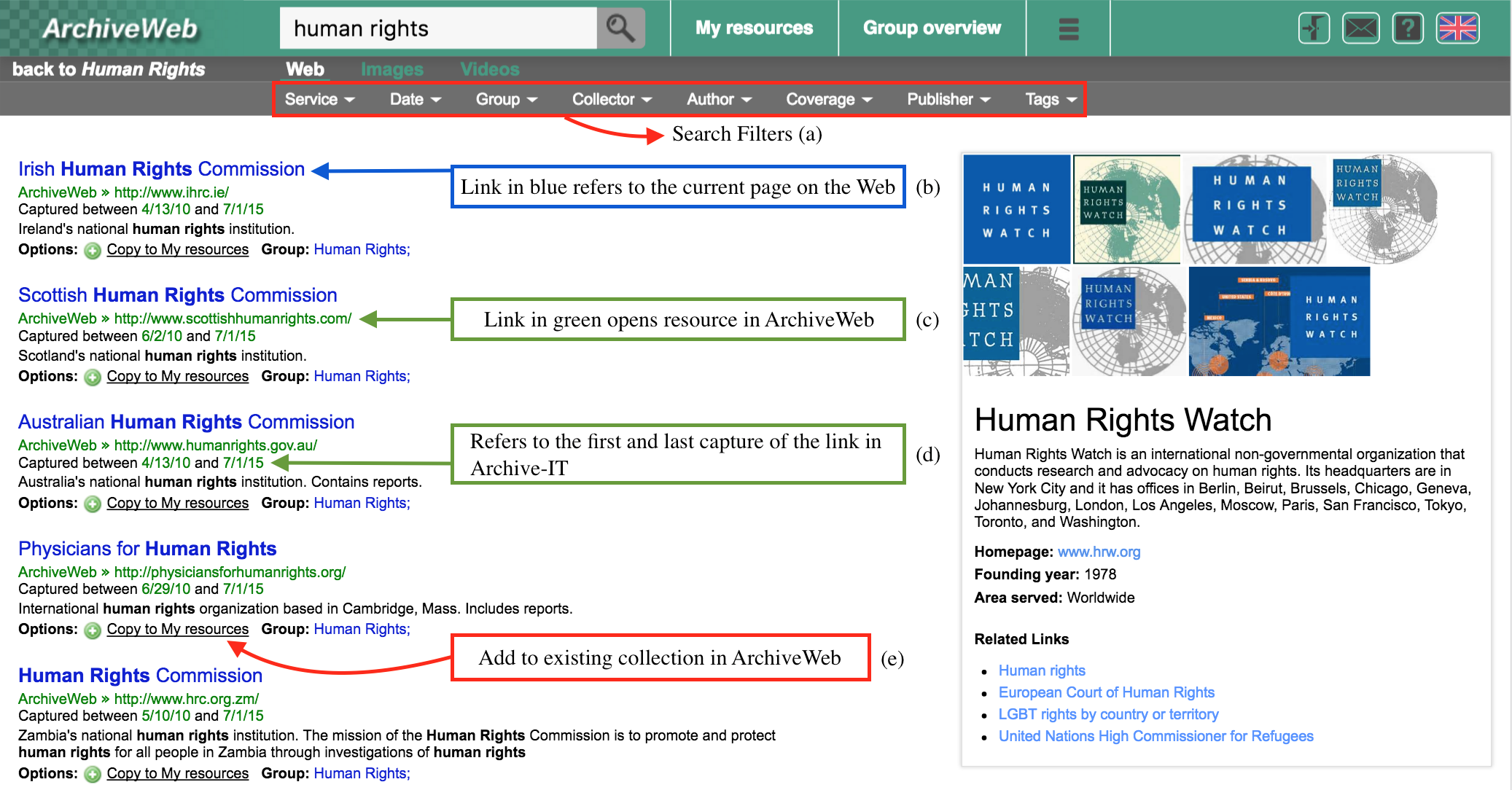}
\caption{Search Page}
\label{search}
\end{figure*}

ArchiveWeb provides a keyword-based search system that returns results
from ArchiveWeb collections as well as from the live web
(Fig. \ref{search}) using the Bing Search
API.\footnote{\url{http://datamarket.azure.com/dataset/bing/search}} ArchiveWeb
collections includes Archive-It collections as well as user created
collections stored in the system. If users search for a keyword (e.g.,
``human rights''), ArchiveWeb returns a list of results from these
sources, indicating whether the resource comes from a specific
Archive-It/ArchiveWeb collection or from the live web
(Fig. \ref{search}b). The search results are presented in this manner,
so that live web results which are not yet part of any collection
could be added to existing ones.  The results being returned from
Archive-It collections are based on a full metadata index for these
collections. A full-text index of all content is planned for the next
version. Besides webpages, images (from Bing, Flickr, Ipernity) and
videos (from YouTube, TED, Yovisto, Vimeo) can also be searched, but
not yet from Archive-It collections.  For Archive-It collections, each
search result provides a pointer to the resource item saved in
ArchiveWeb (Fig. \ref{search}c) and displays information about when it
was captured in Archive-It (Fig. \ref{search}d).  Live web results are
not yet saved in ArchiveWeb but can be copied by the user into a group
(Fig. \ref{search}e); in this case, the Options dialog box
(Fig. \ref{search options}) provides details about when/if this page
has been indexed by the Internet Archive, retrieving the information
from the Wayback CDX server
API.\footnote{\url{https://github.com/internetarchive/wayback/tree/master/wayback-cdx-server}}

\begin{figure*}[!ht]
\includegraphics[width=\textwidth]{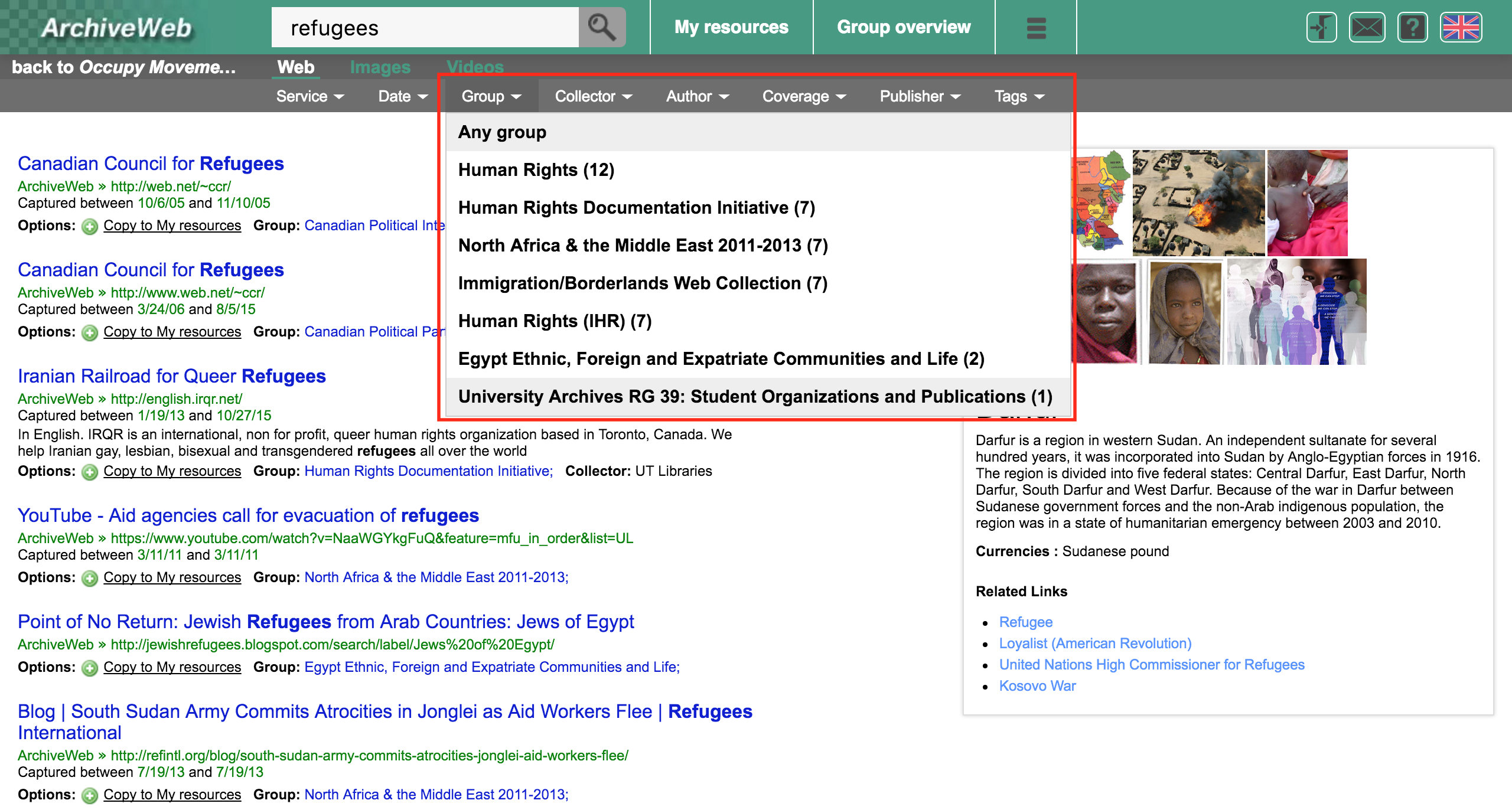}
\caption{Refinement through Faceted Search Filters}
\label{facet}
\end{figure*}

Search results can be refined using faceted search filters visible
below the search box (Fig. \ref{facet}). For example, for our
evaluation it was important to provide filters to show only the
resources in a specific collection (Filter: \textit{Group}) or that
were archived by a specific institution (Filter: \textit{Collector}).

\subsection{Organization}\label{sec:group} 

ArchiveWeb provides the functionality to organize resources into
collections (groups of resources) according to clearly defined and
coherent themes/topics. This functionality allows working with
existing groups, creating new collections/groups and sub-groups,
adding new resources to groups, and moving resources between groups.

\begin{figure*}[!ht]
\includegraphics[width=\textwidth]{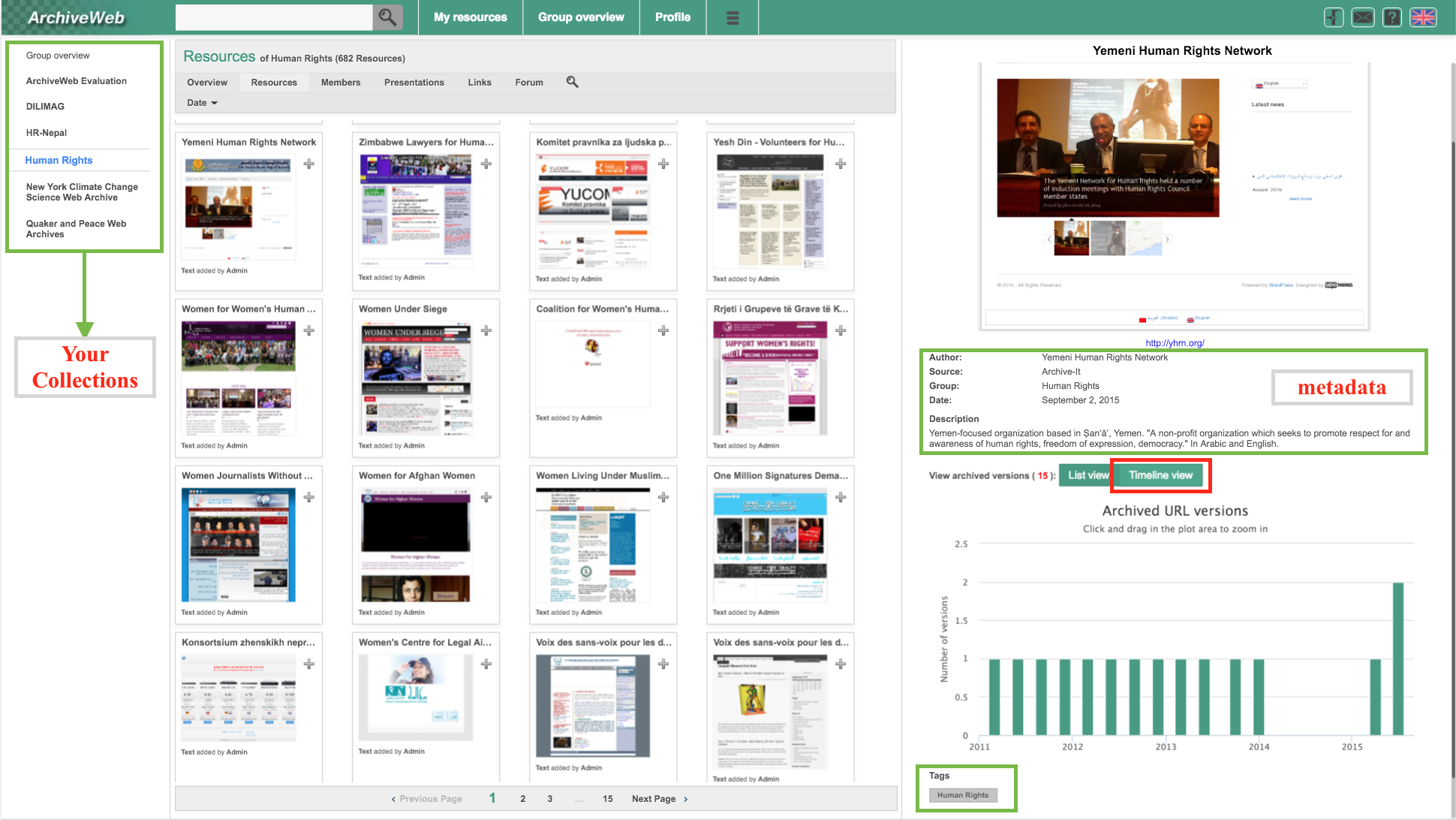}
\caption{Group Resources Interface}
\label{group}
\end{figure*}

A group overview interface allows
browsing through existing collections available within ArchiveWeb and
the collections that a user has created or joined. Descriptions for
every collection provide information about the topic/theme and what
kind of resources it contains. For Archive-It collections, the
description reports the details available on the public Archive-It
interface along with a reference link to the specific collection in
the Archive-It platform. It is possible to filter collections by
searching through the \textit{title} and \textit{description} fields.

After joining specific groups/collections users have the opportunity
to edit metadata of existing resources and to contribute new resources
from other existing collections or from the live web. The ArchiveWeb
collections, derived directly from their Archive-It counterparts, are
``read only''.  Users can browse through the resources of each
collection using the advanced visualization and exploration
functionalities of ArchiveWeb, but they cannot change the original
collection. They can create a copy of the entire ``read-only'' group
or add individual resources to their own collections by selecting
resources individually, as well as merge multiple existing collections
into one new collection.

Users can also organize their resources into sub-groups within
collections in order to group resources that are related to a similar
subtopic. Resources can be uploaded to a collection either from the
desktop or by suggesting a URL. Each group has a specific interface
that visualizes all thumbnails of the most recent snapshot of each
resource when it was added to the group (Fig. \ref{group}). Resources
that no longer exist on the web, or do not have a redirect available,
display a thumbnail with the message ``The page is no longer available
on the web'' (we are currently implementing the functionality to
upload a thumbnail of an earlier capture of the page taken from the
Internet Archive Wayback Machine). The overview interface within a
group provides a summary of the activities of various members of the
group, including the actions of newly added resources, resources which
were edited or deleted, and users who have joined/left the group.

\subsection{Collaboration}\label{sec:enriching}

After a resource is added to a ArchiveWeb collection, the resource can
be enriched with additional comments and tags (in addition to the
metadata already provided). The comments on a resource can be used to
discuss why this resource has been chosen as a seed for a collection,
to decide upon the crawl frequency and crawl depth, as well as any
other issues and discussions which should be documented during the
collection building process (Fig. \ref{annotation}). By exchanging
comments, collaborators can also discuss and decide on the relevance
of a suggested seed resource for a particular
collection.\footnote{ArchiveWeb is not (yet) directly coupled to the
  Archive-It administrative interface: curators have to switch to this
  interface to add suggested seed URLs to the corresponding Archive-It
  collection.} The use of tags helps categorize or label a resource
with subjects or topics covered by the resource, making it possible to
browse collections by filtering based on certain tags. Users can also
edit metadata such as title, description, and author fields.

During the collection building process, multiple collaborators can
join a newly created ArchiveWeb collection and suggest seed URLs from
the live web or from existing collections in ArchiveWeb.
Collaborators can add comments and tags to these seed URLs as
highlighted above to indicate the importance of this resource to the
collection. Once the collaborators have finalized the seed URLs, they
use the Archive-It system to curate this collection by specifying the
seeds with corresponding metadata, crawl depth, and frequency. In the
future, we will investigate a direct connection between ArchiveWeb and
the Archive-It system, in order to provide a seamless environment for
working with and archiving Web collections.

The system allows users to archive a single page or a website
(resource) by clicking on the ``\textit{Archive Now}'' button which
sends a request to the Wayback Machine to archive it. This
functionality is similar to the ``\textit{Save Page Now}'' feature of
the Wayback machine, and it supports users to gather captures easily
as they work within the system. All captures from Archive-It as well
as the Wayback captures in ArchiveWeb added using the
``\textit{Archive Now}'' functionality are visualized both as a
\textit{list} and in a \textit{timeline} view to help users navigate
through the different archived versions that are available. The \textit{timeline} view visualizes a timeline displaying the number of archived versions aggregated by month for each resource. Figure
\ref{group} shows a timeline view on the bottom right.

\section{Evaluation and Feedback 2016}\label{sec:Evaluation}

At the end of 2015, the release of ArchiveWeb, which took into
account the initial requirements collected in summer 2015, was ready
for a first evaluation. In February 2016, we invited all experts, who
provided the initial user requirements, to participate in a task-based
evaluation of the system. All of them agreed to participate.

For the evaluation, we imported 200 publicly accessible web archive
collections from Archive-It into ArchiveWeb in order to test the
potential of the system to support collaborative work with such
collections. The evaluation was done using a task-based evaluation
design, analyzing both quantitative interaction data (usage logs)
 and qualitative feedback in the form of interviews. We asked
the experts to carry out two sets of tasks: \textit{individual} and
\textit{collaborative}. The \textit{individual tasks} included
creating a sub-collection from an existing collection of their
institution, enriching an existing collection with additional relevant
resources from the live web and annotating resources with information
about why they should be included and how often they should be
archived. The \textit{collaborative tasks} involved selecting ten
featured resources from their collections, including them into a joint
group shared with all other evaluators, and discussing the reasons why
such resources were included. The final task involved creating a new
collection about a shared topic among the evaluators, searching for
relevant seed URLs for this collection using ArchiveWeb, and agreeing
with the rest of the evaluators about which seeds should be included,
what should be the archiving frequency, and why.

The evaluation period spanned about three weeks, after which we
invited the evaluators to give us feedback through a questionnaire and
through follow-up interviews to fully understand their experience. In
the following, we will summarize the feedback we received and point out
specific interesting responses and suggestions, based on the questions
we asked. The headings denote the questions being asked.

\subsection{Usage Log Analysis}

In the ArchiveWeb system, we log the various actions carried out by
the user while working with the system. The types of actions logged
include annotation (tags or comments added to resources), searching
(queries issued, type of search such as image, video, webpages, search filters used),
resource-level actions (additions/deletions/edits) and group-level
actions (metadata edits/subfolders actions/search within a group/group
merging).

After the evaluation, we analyzed the usage logs to understand how the
evaluators worked with the ArchiveWeb system and how the various tasks
were completed. For the {\em initial collaborative task} of selecting
featured resources from their collections to be added to a shared
group with the evaluators, 6 of the 8 evaluators participated in this
task. Most of the evaluators added few featured resources from their
own collections to the shared group, but one of the evaluators added
resources directly from Bing by searching for known URLs, and one
evaluator added resources that were previously uploaded to a newly
created collection. All of the evaluators tagged their resources, and
one evaluator added comments to discuss why a resource was
included. None of the evaluators added comments on resources added by
others; thus there was no discussion or collaboration between the
evaluators. This could be due to different interests in topics among
the evaluators. Evidence from the usage logs shows that the evaluators
could not carry out the {\em final collaborative task} of creating a
new shared collection based on a shared topic and collecting seed URLs
due to the short time period of the evaluation and limited
communication between the evaluators.

As specific examples, let us here summarize the {\em individual tasks}
carried out by two evaluators as observed from the logs. A Web archiving coordinator of a university library initially joined his institution's various collections covering topics such as Human Rights, Arts \& Humanities, Society \&
Culture. He then created a new collection by merging resources
from two existing collections. Later, he carried out 20 searches related
to climate justice or urban planning with the use of filters such as
``author'', ``tags'', ``groups''. He added tags and comments related to
crawling strategies of resources such as ``one-time
capture''. Finally, he worked for 50 min on the initial
collaborative task and added 10 featured resources to the joint group
with comments discussing their importance.

A Web archiving coordinator of a national museum created a new collection and added about 10
resources from an existing collection of her institution. Then she carried out 5 searches related to photography/ photography
galleries using filters such as ``service'' or ``groups'' and added 5
resources from Bing to a sub-folder created within the new
collection. Later, she added tags and comments such as ``photo gallery''
and ``archive monthly''. Finally, she searched for 5 known
resources from Bing and added them along with 3 other resources from
an existing collection to the joint group with tags, in order to get a
feel of how close the workflow of ArchiveWeb was to the existing
one. This final task was completed in 20 min.

\subsection{Evaluation: Questions and Answers}

As we mentioned in Sect. \ref{sec:User Requirements}, one of the
goals of our work is to contribute further improvements for working
with web archive collections, especially regarding collaborative
search and exploration of resources, as well as organizational
functionalities. For this reason, we asked the evaluators to give us
feedback about the current functionalities, as well as suggestions for
further improvement. In general, all evaluators appreciated the
curatorial functionalities provided by the system. The answers were
analyzed manually and are categorized according to the four main steps
reported in the ArchiveWeb User Model in Fig. \ref{aum}.

\subsubsection{Nomination}

\paragraph{Seed URL Discovery.} The integration of searching across
web archive collections as well as the live web, providing the ability
to suggest additional material that could be archived, is a new
feature which is not present in existing systems used for curation/creation of web archive collections, and it was much appreciated by
the evaluators: {\em I really liked the feature of searching the web
  in conjunction with the web archive collections available in the
  system, an entirely new feature which was seamless and
  straightforward.}

Others positively stressed the possibility to archive web pages
directly from the interface: {\em I also like that the ``Archive Now''
  feature is so well embedded in the platform - it will make it very
  easy for people to gather new captures as they work, without having
  to leave the interface}.

\paragraph{Nomination tool.} Nomination for collaborative collection
building was mentioned as well: {\em A nomination function would be
  nice -- like URL suggestion instead of having to upload all
  resources. ... The inclusion of screenshots and ability to pull in
  resources from live web could make this tool a better option than
  the UNT nomination
  tool~\footnote{\url{http://digital2.library.unt.edu/nomination/}}
  for collaborative collection building. ... Ability for non-members
  to submit tags for content or suggest content for future archiving
  could be submitted for approval first.}

\subsubsection{Search}

A few remarks concerned our {\em ranking of results} and included
suggestions for (i) providing a transparent explanation of how the
relevance ranking is determined, (ii) providing the total number of
captures of a page along with the capturing period in order to
determine the popularity of an archived result, and (iii) cues as to
how many results were returned. Users in general liked the information
provided on the results page: {\em Yes, I think the highlighted search
  term and snippet are sufficient for each item in search results. I
  like that the web archive hits are placed before the live web
  hits}, but others also mentioned: {\em The results would provide
  enough information if the archived resources had rich enough
  metadata. It would make way more sense if there were more tags}.

A related suggestion concerned results from different sources. Users
liked the possibility to retrieve results from several sources: {\em
  The search results seem to offer enough information. I appreciate
  that the user can filter by the source of the content (e.g.,
  Archive-It, ArchiveWeb, Bing)}, but others asked for more clearly
separating results from different sources or defaulting to particular
sources: {\em I do wonder about the utility of having the Bing results
  mixed with the Archive-It results unless the idea is that the user
  will use these results to add resources to collection ... I would
  think that the default search results would just be Archive-It
  collections then with the opportunity to expand out next.}

Other remarks asked for advanced search functionalities such as limit
to collection, limit to domain/path, search within title only. The experts also requested full-text search of Archive-It records
(\textit{WARCs}) instead of just metadata search.


\subsubsection {Organization}

\paragraph{Efficiency through the interface.}

From conversations with Archive-It partners and the NDSA Web Archiving
Survey, it became clear that web archiving on the whole takes a lot of
time to do well. Therefore, for curatorial needs, efficiency is a
critical issue. This asks for making it as quick and intuitive as
possible to move from one task to another and making tasks as
efficient as possible (providing bulk operations, summary views, etc.)

One of the evaluators stated that: {\em ArchiveWeb increases
  curatorial functionality over that of Archive-It, e.g.,  on-the-fly
  creation of groups, moving resources between groups, and easily
  annotating resources}.


\paragraph{Collection management.} Several of the new functionalities
of ArchiveWeb were positively highlighted by the evaluators including
(i) the ability to curate new arbitrary collections of seed records
from across multiple web archive collections, (ii) the possibility of
having resources exist at multiple levels (e.g., personal, group,
sub-folder), and (iii) the ability to create collaborative collections
with colleagues from various institutions.

\paragraph{Improving the user interface.} A few remarks were
specifically referring to curatorial functionalities, such as bulk
operations: selection of multiple resources either by clicking,
searching, or filtering and then adding/moving to a collection;
collective tagging of multiple seed records; and initiating archiving for
multiple seed records. One of the evaluators mentioned: {\em It'd be
  great to tag similar collections and allow them to be grouped
  together without having to create a new collection.}

An \textit{undo} feature would also be useful for ArchiveWeb: {\em I
  would like to see some sort of ``undo'' feature for mistakes that
  are made in the system (for example, if I deleted something from a
  group and immediately felt it was in error, I could undo that
  action}).

\paragraph{Exporting resources and metadata.} Several experts missed a
way for exporting resources and metadata for research / usage outside
of ArchiveWeb, and means to request derivative data from Archive
Research Services such as
WATs\footnote{\href{https://webarchive.jira.com/wiki/display/ARS/WAT+Overview+and+Technical+Details}{Web
    Archive Transformation} (WAT) files contain key metadata such as
  capture information, essential text and link data, and other
  information extracted from (W)ARC files.},
WANEs\footnote{\href{https://webarchive.jira.com/wiki/display/ARS/WANE+Overview+and+Technical+Details}{Web
    Archive Named Entities} (WANE) files contain a list of people,
  places and organizations mentioned in each valid archived record,
  extracted using Stanford Named Entity Recognizer.},
LGAs\footnote{\href{https://webarchive.jira.com/wiki/display/ARS/LGA+Overview+and+Technical+Details}{Longitudinal
    Graph Analysis} (LGA) files feature a complete list of what URLs
  link to what URLs, along with a timestamp, within an entire web
  archive collection.}, etc.: {\em I would appreciate means of
  interacting with content instead of just searching for web pages,
  e.g., ngrams, word frequency like UK Web Archives
  SHINE~\footnote{\url{https://www.webarchive.org.uk/shine}}}.

Experts liked the ability of our interface to provide an easy overview
over resources and collections using screenshots, but also suggested
to provide {\em screenshots of different captures} and select a
specific capture as default screenshot (the default we implemented for
the evaluation was the screenshot of the last available page).

\subsubsection {Collaboration}

\begin{figure}[!ht]
\centering
\includegraphics[width=\columnwidth]{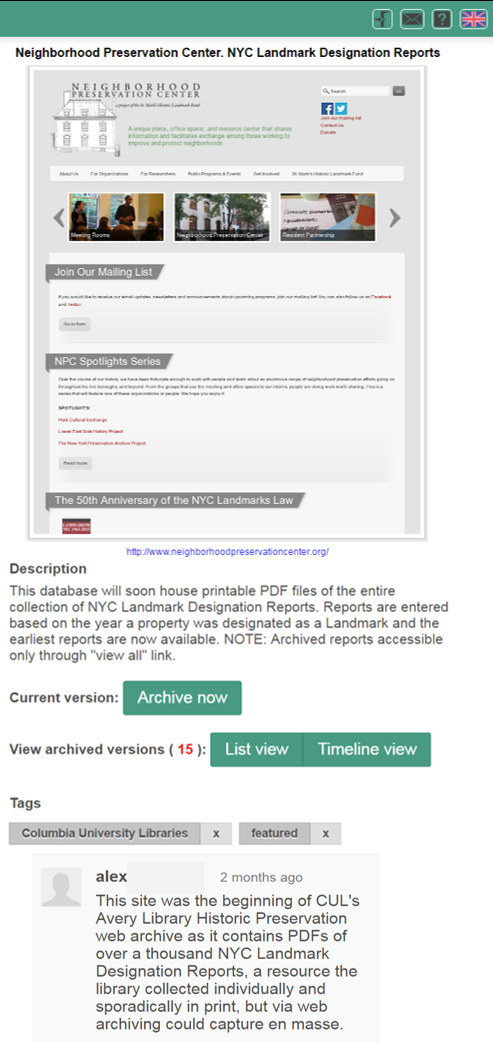}
\caption{Resource annotation}
\label{annotation}
\end{figure}
The evaluators positively highlighted the usefulness of ArchiveWeb to
facilitate collaborative collection building and to make web archives
more useful to researchers: (i) {\em It is a really fun tool to use!
  ... I enjoyed using ArchiveWeb and think it has great potential in
  facilitating collaborative endeavors and potentially collection
  building}, (ii) {\em I think it's incredibly easy to
  use. Institutions, particularly those without a great deal of IT
  support, would really benefit from having an interface like this to
  help entice users to interact with web archive collections. One of
  the most exciting aspects of this project is the potential for users
  to help us build and enrich collections through suggesting resources
  and annotating what we've already archived. I'm excited to see how
  this project progresses}.

\paragraph{Collection Enrichment.} One of the new functionalities that
was positively highlighted by the evaluators is the possibility to add
schema-agnostic classification and other information using tags.  One
participant (Technical Manager) asked for more metadata fields for
newly added resources besides \textit{title}, \textit{author} and
\textit{description}, to make it easier for curators from different
institutions to collaboratively build collections: {\em Tagging
  functionality is helpful, but not quite same as being able to define
  a custom field name and supply a value for that field.}

Other appreciated functionalities were (i) the ability to use
notes/comments in order to highlight the motivations for collecting
(Fig. \ref{annotation}): {\em I also have not previously been able to
  tag or annotate resources within collections (Archive-It does allow
  for adding notes in their new release, but these are only useful as
  an administrator of an account and not to the public/user)}, and
(ii) the display of the capture frequency, highly useful when
developing collaborative collections across institutions.



Regarding annotations, suggestions included auto-suggestion of tags
from within the same collection, optionally allowing tag suggestions
from controlled vocabularies, and conducting more robust mining of the
text of archived resources to suggest possible annotations, plus the
possibility to include external contributors for tagging: {\em It'd be
  great to turn on external tagging, or pick certain trusted external
  users to tag.}

Finally, we asked for further suggestions on any topic important to
the users.

Many evaluators found the system rather intuitive, but had to refer to the
documentation about the functionalities that we had provided in order
to feel comfortable using the system: (i) {\em The PDF instructions
  helped since there are not a lot of ``helpers'' or ``tool tips'' in
  the interface itself. More help features baked into the interface
  would help as not all the features are well defined or obvious}.
(ii) {\em Additional text could be useful to users - I would suggest
  hover text, where the user holds the mouse over something and a
  description would pop up, with an option to close it and not see it
  again, if desired}. Also, having information about the model
underlying the metadata (schema, ontology) could help understand how
this metadata could be shared with other systems.

One of the experts suggested that for users who are unfamiliar with
web archives, it would be good to provide some information highlighting
the need and challenges of web archives: {\em I think users unfamiliar
  with web archives might need some tools or modules or something to
  help better understand some of the challenges of web
  archives. Otherwise, they may just think of web archives as a
  ``blackbox'' and accept collections uncritically, unaware of why
  some things work and some things don't. That could just be something
  like a mouse over or an FAQ entry though}.

Table \ref{table:2} summarizes the results of the evaluation, with
particular attention on the phases supported by the ArchiveWeb User
Model.

In the next iteration of the evaluation, we will focus specifically on
collaborative tasks, in order to clearly understand the collaborative
potential of the system for Web archiving.  The planned evaluation
will span 1-2 months in order to give our partners more time to
collaboratively discuss and build a collection together.

\begin{table*}[htbp]
\centering
\begin{tabularx}{\textwidth}{|X|X|}
\hline
Current Functionalities & Required Functionalities \\
\hline
\multicolumn{2}{|X|}{Nomination} \\
\hline
\begin{itemize}[leftmargin=*]
\item Seed URL Discovery: ability to curate new arbitrary collections
  of seed records from across multiple web archive collections 
\item Possibility to archive web pages directly from the interface
\end{itemize} &
\begin{itemize}[leftmargin=*]
\item A ``nomination'' function would be nice - like URL suggestion -
  instead of having to upload all resources 
\item Display the capture frequency
\end{itemize} \\
\hline
\multicolumn{2}{|X|}{Search} \\
\hline
\begin{itemize}[leftmargin=*]
\item Combined searching of seed metadata with search engine of live web
\item Dynamic effects of ``List View'' and ``Timeline View''
  visualizations for individual seed records 
\end{itemize} &
\begin{itemize}[leftmargin=*]
\item Full text search of WARC
\item Google as well as Bing as web search options
\item Advanced search options (e.g., limit to collection, limit to
  domain/path, search within title only) 
\item Provide more details about the total number of captures of a
  page along with the capturing period 
\item Means of interacting with content instead of just searching for
  web pages (e.g., ngrams, word frequency) 
\end{itemize} \\
\hline
\multicolumn{2}{|X|}{Organization} \\
\hline
\begin{itemize}[leftmargin=*]
\item On-the-fly creation of groups (sub-collections)
\item Possibility of having resources exist at multiple levels (e.g.,
  personal, group, subfolder) 
\end{itemize} &
\begin{itemize}[leftmargin=*]
\item Bulk operations: select multiple resources and work with all at
  once (e.g., copy, move, edit, group, delete, annotate, share) 
\item Exporting of resources and metadata for research/usage outside
  of ArchiveWeb 
\item Provide information about the model underlying the metadata
\end{itemize} \\
\hline
\multicolumn{2}{|X|}{Collaboration} \\
\hline
\begin{itemize}[leftmargin=*]
\item Seed collection: ability to share seed URLs across collections
  with colleagues from various institutions 
\item Annotation: ability to use notes/comments in order to
  highlight the motivations for collecting 
\item Forum for users to discuss collections
\end{itemize} &
\begin{itemize}[leftmargin=*]
\item Collection Enrichment: possibility to add schema-agnostic
  classification and other information using tags 
\item Request collaborators within the system
\item Allow external contributors for tagging (could be submitted for
  approval first) 
\end{itemize} \\
\hline
\end{tabularx}  
\caption{System Evaluation - Summary of results}
\label{table:2}
\end{table*}

\section{Future Improvements in the Next Release}\label{sec:Future
  Improvements}

Based on the suggestions provided by the evaluators, we plan to
incorporate the following new functionalities in the next release, in
addition to several smaller user interface improvements asked for
during the evaluation:

\paragraph{Thumbnail Visualization.} We are working on a specific
  visualization to display the thumbnails for different captures of an
  individual resource/seed, where the curator/user can choose how
  many different thumbnails he wants to view.
\paragraph{Search Result Aggregation.} We are also working on a new
  search interface design that aggregates results from different
  sources, displaying some representative examples for each source,
  and in addition will allow users to select a default source for
  which we display the results in detail, i.e., using a list of results
  view similar to the current design.
\paragraph{Bulk Operations.} We will also work on bulk operations which
  allow users to select multiple resources by clicking, searching, or
  filtering and then adding them to a collection. The system will
  also support bulk editing of resources within a collection such as
  collective tagging (e.g., to assign a tag to multiple seed records),
  as well as tagging similar resources across multiple collections and
  then allowing the option to group them together into a new
  collection.
\paragraph{Exporting Resources.} We will also implement the ability
  to export resources and metadata from collections for research or
  usage outside of ArchiveWeb, for registered users who are allowed to
  use this export functionality (usage will have to be restricted
  because of potential copyright issues).
\paragraph{Advanced Search.} We will extend our advanced
  search and faceted search options to limit search within certain
  collections, specific domains, or paths, and to search within title
  or description only.
\paragraph{Fulltext Search.} We will also incorporate full-text search
  of Archive-It websites instead of just the metadata, even though not
  all evaluators even realized that this functionality was not yet
  provided, as rich metadata descriptions are available for many
  collections. Obviously, this fulltext search capability will have to
  go along with appropriate ranking optimizations in order to take
  into account the longitudinal character of Archive-It collections,
  which usually include many captures per resource which have to be
  considered.
  \paragraph{Collaborative Aspects.} We will allow external
  contributors to suggest tags, which can then be approved by the
  curators of the collection.  The system will also support the
  possibility to invite collaborators to join collections from within
  ArchiveWeb.

\section{Conclusions}\label{sec:Conclusion}

In this paper, we discussed the ArchiveWeb system which supports
collaborative exploration of web archive collections. We provided a
description of the main features of the system such as (i) searching
across multiple collections as well as the live web, (ii) grouping of
resources for creating new collections or merging existing ones, as
well as (iii) collaborative enrichment of resources using comments and
tags.

The system has been developed based on an iterative evaluation-driven
design-based research approach. Starting from a platform which already
supported collaborative search and sharing of web resources,
ArchiveWeb was designed to address web archive expert users'
requirements (librarians and curators in archiving institutions), as
described in Sect. \ref{sec:User Requirements}. The resulting
ArchiveWeb system, fully functional and publicly available,\footnote{\url{http://archiveweb.l3s.uni-hannover.de/aw/index.jsf}} has
been evaluated through a task-based evaluation study carried out with
the same experts who participated in the preliminary
investigation. After the quantitative analysis of the logs and the
qualitative feedback from the evaluation, we are now incorporating new
features such as exporting resources, bulk editing operations, new
visualizations, and advanced search, which will be available in the next
release of the system.

\begin{acknowledgements}\label{sec:Acknowledgements}
  We especially thank Jefferson Bailey from the Internet Archive who
  provided us with the contacts to his colleagues at university
  libraries and archiving institutions, and for his helpful comments
  during the requirements and evaluation phase. We are also grateful
  to all experts, who participated with enthusiasm in our evaluation,
  providing valuable feedback and useful suggestions to improve the
  ArchiveWeb system. This work was partially funded by the European
  Commission in the context of the Alexandria project (ERC advanced
  Grant No. 339233).
\end{acknowledgements}

\bibliographystyle{spmpsci}      
\bibliography{ArchiveWeb_IJDL}

\begin{thebibliography}{10}
\providecommand{\url}[1]{{#1}}
\providecommand{\urlprefix}{URL }
\expandafter\ifx\csname urlstyle\endcsname\relax
  \providecommand{\doi}[1]{DOI~\discretionary{}{}{}#1}\else
  \providecommand{\doi}{DOI~\discretionary{}{}{}\begingroup
  \urlstyle{rm}\Url}\fi

\bibitem{twaw11}
Alonso, O., Str{\"o}tgen, J., Baeza-Yates, R., Gertz, M.: Temporal information
  retrieval: challenges and opportunities.
\newblock In: Proceedings of the 1st international temporal web analytics
  workshop (TWAW 2011) associated to WWW'11, pp. 1--8 (2011)

\bibitem{walcm13}
Bragg, M., Hanna, K., Donovan, L., Hukill, G., Peterson, A.: The {Web}
  {Archiving} {Life} {Cycle} {Model}.
\newblock White Paper  (2013).
\newblock Available at
  \url{http://ait.blog.archive.org/files/2014/04/archiveit_life_cycle_model.pdf}

\bibitem{cutrell06}
Cutrell, E., Robbins, D., Dumais, S., Sarin, R.: Fast, {Flexible} {Filtering}
  with {Phlat}.
\newblock In: Proceedings of the SIGCHI Conference on Human Factors in
  Computing Systems (CHI '06), pp. 261--270 (2006)

\bibitem{dougherty09}
Dougherty, M., van~den Heuvel, C.: Historical infrastructures for web
  archiving: Annotation of ephemeral collections for researchers and cultural
  heritage institutions (2009).
\newblock Available at
  http://web.mit.edu/comm-forum/mit6/papers/Dougherty\_Heuvel.pdf

\bibitem{dumais03}
Dumais, S., Cutrell, E., Cadiz, J., Jancke, G., Sarin, R., Robbins, D.C.: Stuff
  {I've} seen: a system for personal information retrieval and re-use.
\newblock In: Proceedings of the 26th Annual International ACM SIGIR Conference
  on Research and Development in Informaion Retrieval (SIGIR '03), pp. 72--79
  (2003)

\bibitem{Fernando2016}
Fernando, Z.T., Marenzi, I., Nejdl, W., Kalyani, R.: Archiveweb:
  Collaboratively {Extending} and {Exploring} {Web} {Archive} {Collections}.
\newblock In: Proceedings of 20th International Conference on Theory and
  Practice of Digital Libraries: Research and Advanced Technology for Digital
  Libraries (TPDL '16), pp. 107--118 (2016)

\bibitem{gomes11}
Gomes, D., Miranda, J., Costa, M.: A survey on web archiving initiatives.
\newblock In: Proceedings of the 15th International Conference on Theory and
  Practice of Digital Libraries: Research and Advanced Technology for Digital
  Libraries (TPDL '11), pp. 1045--1050 (2011)

\bibitem{jackson16}
Jackson, A., Lin, J., Milligan, I., Ruest, N.: Desiderata for {Exploratory}
  {Search} {Interfaces} to {Web} {Archives} in {Support} of {Scholarly}
  {Activities}.
\newblock In: Proceedings of the 16th Joint Conference on Digital Libraries,
  JCDL '16, pp. 103--106 (2016)

\bibitem{lieser09}
Lieser, W.: Digital {Art} ({Art} {Pocket}).
\newblock H.F.Ullmann Publishing GmbH (2009)

\bibitem{lin14}
Lin, J., Gholami, M., Rao, J.: Infrastructure for supporting exploration and
  discovery in web archives.
\newblock In: Proceedings of the 23rd International Conference on World Wide
  Web (WWW '14), pp. 851--856 (2014)

\bibitem{marenzi14}
Marenzi, I.: Multiliteracies and e-learning2.0.
\newblock In: G.~Blell, R.~Kupetz (eds.) Foreign Language Pedagogy, Content and
  Learner Oriented, vol.~28. Peter Lang (2014)

\bibitem{marenzi12}
Marenzi, I., Nejdl, W.: I search therefore I learn - Active and collaborative
  learning in language teaching: Two case studies, pp. 103--125.
\newblock Collaborative Learning 2.0: Open Educational Resources. IGI Global
  (2012)

\bibitem{tlt12}
Marenzi, I., Zerr, S.: Multiliteracies and active learning in {CLIL} - the
  development of {LearnWeb2.0}.
\newblock In: IEEE Transactions on Learning Technologies (TLT) (2012)

\bibitem{odijk15}
Odijk, D., G\^{a}rbacea, C., Schoegje, T., Hollink, L., de~Boer, V., Ribbens,
  K., van Ossenbruggen, J.: Supporting {Exploration} of {Historical}
  {Perspectives across Collections}.
\newblock In: Proceedings of 19th International Conference on Theory and
  Practice of Digital Libraries (TPDL '15), pp. 238--251 (2015)

\bibitem{padia12}
Padia, K., AlNoamany, Y., Weigle, M.C.: Visualizing {Digital} {Collections} at
  {Archive-It}.
\newblock In: Proceedings of the 12th ACM/IEEE-CS Joint Conference on Digital
  Libraries (JCDL '12), pp. 15--18 (2012)

\bibitem{archiving07}
Ras, M., van Bussel, S.: Web archiving user survey.
\newblock Technical report, National Library of the Netherlands (Koninklijke
  Bibliotheek)  (2007)

\bibitem{stalker05}
Stalker, P.J.: Gaming {In} {Art}: {A} {Case} {Study} {Of} {Two} {Examples} {Of}
  {The} {Artistic} {Appropriation} {Of} {Computer} {Games} and {The} {Mapping}
  {Of} {Historical} {Trajectories} {Of} {'Art Games'} {Versus} {Mainstream}
  {Computer Games}.
\newblock University of Witwatersrand  (2005)

\bibitem{wtst11}
Weikum, G., Ntarmos, N., Spaniol, M., Triantafillou, P., Bencz{\'u}r, A.,
  Kirkpatrick, S., Rigaux, P., Williamson, M.: Longitudinal {A}nalytics on
  {W}eb {A}rchive {D}ata: {I}t's {A}bout {T}ime!
\newblock In: Proceedings of the $5^{th}$ biennial {C}onference on {I}nnovative
  {D}ata {S}ystems {R}esearch ({CIDR}), {A}silomar, {CA}, {USA}, {J}anuary
  9-12, pp. 199--202 (2011)

\bibitem{jane17}
Winters, J.: Tackling complexity in humanities big data: {From} parliamentary
  proceedings to the archived web.
\newblock In Turo Hiltunen, Joe McVeigh and Tanja S{\"a}ily (eds.), Big and
  Rich Data in English Corpus Linguistics: Methods and Explorations. Studies in
  Variation, Contacts and Change in English. Helsinki: VARIENG. (Forthcoming
  2017)

\end{thebibliography}
\end{document}